\documentclass{article}

\usepackage[english]{babel}

\usepackage{graphicx}
\usepackage{indentfirst}
\usepackage{float}
\usepackage{subfigure}
\usepackage{color}
\usepackage{tabularx}
\usepackage{amsmath}      
\usepackage{amsfonts}     
\usepackage{mathrsfs}
\usepackage{ulem}    

\usepackage{multirow}
\usepackage{makecell}
\usepackage{booktabs}
\usepackage{geometry}
\usepackage{algpseudocode}
\usepackage{tcolorbox}
\usepackage{tabu,makecell,multirow}
\usepackage{hyperref}

\title{Nash equilibrium selection by eigenvalue control}
\author{ Wang Zhijian \\ Experimental social science laboratory, Zhejiang University, China}

\begin{document}
\maketitle

\begin{abstract}
People choose their strategies through a trial-and-error learning process in which they gradually discover that some strategies work better than others. The process can be modelled as an evolutionary game dynamics system, which may be controllable. In modern control theory, eigenvalue (pole) assignment is a basic approach to designing a full-state feedback controller, which can influence the outcome of a game. This study shows that, in a game with two Nash equilibria, the long-running strategy distribution can be controlled by pole assignment. We illustrate a theoretical workflow to design and evaluate the controller. To our knowledge, this is the first realisation of the control of equilibrium selection by design in the game dynamics theory paradigm. We hope the controller can be verified in a laboratory human subject game experiment.
\end{abstract}

\tableofcontents

%
%
%
%
%
%
%
%
%
%
%
%
%
%
%
%
%
%
%
%
%
%



\section{Introduction}

\paragraph{Motivation}
Game theory study the strategy interaction between intelligent species, e.g., human.
Its statics paradigm predicts equilibrium, and its dynamics paradigm predicts motion. Equilibrium selection is an open question both in the statics \cite{harsanyi1988} and the dynamics \cite{samuelson1998}.

In this study,  using game dynamics paradigm \cite{dan2016,2011Sandholm}, we show how to control game equilibrium selection by exploiting the predictability of the game motion. In another words, instead of statics \cite{harsanyi1988} or existed dynamics \cite{samuelson1998} consideration, we design a mechanism to control the velocity field to influence the equilibrium selection. The controller is designed by the pole assignment approach in modern control theory \cite{pole2017modern,pole2018}.
%
We will provide an overview of the workflow for designing the controller and illustrate it with an example.
\paragraph{Background}

The research goal is primarily based on the two factors listed below:
\begin{itemize}
    \item \textbf{Human strategy behaviour motion can be predicted by the eigensystem of the game dynamics equations. }
    --- Data from game experiments show that human dynamics behaviour is governed by dynamics equations (for example, replicator dynamics) and its linearization at the Nash equilibrium. Previous results have shown that, even in discrete time and discrete strategy games, the linearization approximation works.
    Examples come from the 3-strategy game (the rock-paper-scissors game \cite{dan2014,wang2014social,2015nowak}), the 4-strategy one-population games (\cite{wang2022shujie}), the 5-strategy one-population games  (Yao 2021 \cite{2021Qinmei}), and 4-strategy two-role zero-sum asymmetry game of the O'Neill 1987 game (O'Neill \cite{ONeill1987},
    Wang and Yao 2020 \cite{WY2020}).
These findings demonstrate that the game dynamics theory can accurately capture human subject experimental dynamics behaviours based on distribution, cycle, and converge speed measurements.
    This logic chain can be seen in Figure \ref{fig:workflow2}(a).
    The chain begins with the game and dynamics equations to obtain the rest point (equilibrium), then proceeds to the Jacobian $Jo$, and finally to the eigensystem (which includes eigenvalues, eigenvectors, and eigencycles).
    \item \textbf{A linear dynamics system is controllable}
    --- Applied mathematics shows that the invariant manifold concept provides a clear picture for dynamic process description. This concept has roots in dynamical systems theory\cite{wiki:Dynamical_systems_theory}, a solid branch of mathematics. Dynamical systems deal with the study of the solutions to the equations of motion of systems that are primarily mechanical in nature.
For linear dynamic systems, state-space feedback controller design is a well-studied and applied engineering field, namely modern control theory and application \cite{2015feedback,2010moderncontrol,pole2018}.
\end{itemize}

\paragraph{Main logic}

\begin{table}[h!]
\caption{The game matrix}
\begin{center}
\begin{tabular}{c|rrrrr}
 \hline
&x$_1$&x$_2$&x$_3$&x$_4$&x$_5$\\
 \hline
x$_1$& 0  & 0 &  2 &  0 & -2 \\
x$_2$&2  & 0  & 0  &-2  & 0 \\
x$_3$&0  & 2 &  0 &  2 & -1 \\
x$_4$&$-$2 &  0  & 1 &  0  & 1\\
x$_5$&0  &$-$2 & $-$2 &  1 &  0 \\
 \hline
\end{tabular}
\end{center}
\label{tab:gamemodel}
\end{table}

\begin{figure}
\centering
\includegraphics[width=.5\textwidth]{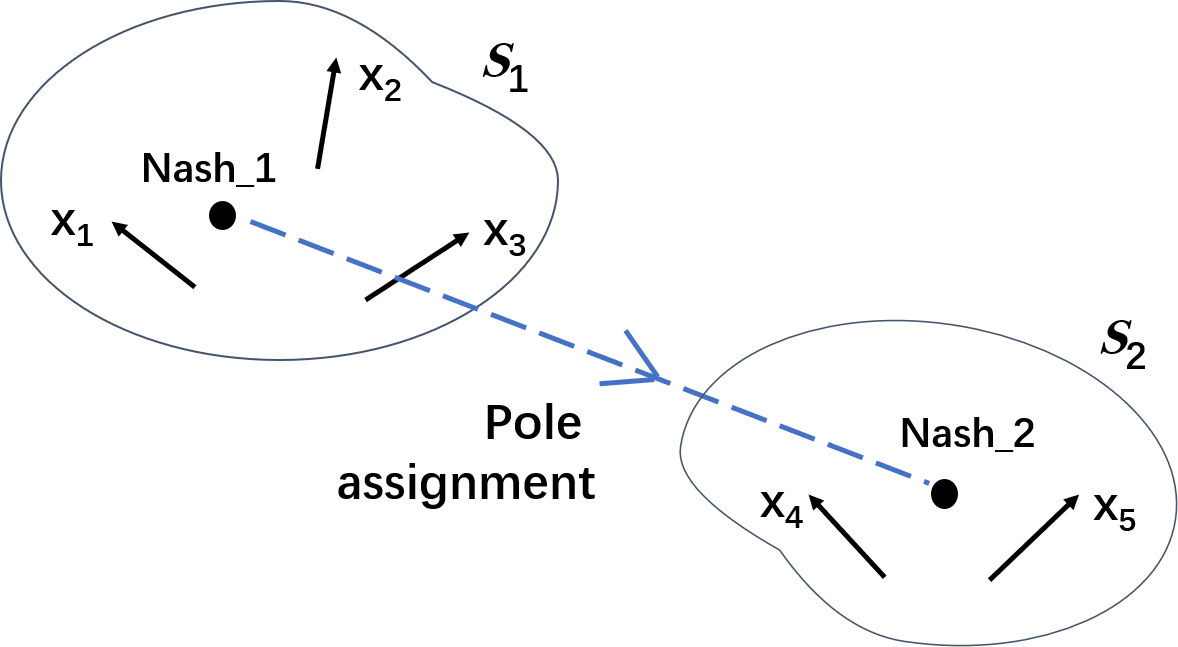}
\caption{Conceptual figure:
In a five strategy game, the two equilibrium (Nash\_1
and Nash\_2) locate in the two sub space $S_1(x_1, x_2, x_3)$
and $S_2(x_4, x_5)$, respectively.
The pole assignment can be designed to make the Nash\_1
being unstable.
As consequence, the long run trajectory will converge to Nash\_2,
which means that the equilibrium Nash\_2 is selected.
\label{fig:concept}}
\end{figure}

As the linear dynamics system can be controlled, we turn to equilibrium selection. We show how to design a mechanism for equilibrium selection by using an example of a symmetric 5-strategy one population game, whose payoff matrix is shown in Table \ref{tab:gamemodel}. The game has and only has two equilibrium (Nash\_1 and Nash\_2) as shown in Fig. \ref{fig:concept}. By using the pole assignment approach, we can control the eigenvalue at equilibrium, equilibrium Nash\_1. In other words, we use pole assignment to control the stability of Nash\_1. As a result, the long-run distribution will remain at Nash\_1 or move to Nash\_2. This is the main logic.

In fact, in the view of game  velocity field:
(1) In game dynamics theory, dynamics equations (e.g., replicator dynamics) describe the velocity field.
(2) The velocity field really exists in the human game experiments of von 1947's two elementary games \cite{von1947}, the standard rock-paper-scissors game and the matching pennies game.
(3) The velocity field can be controlled by the pole assignment approach. Thus, using the pole assignment to influence human dynamics behaviours in a game experiment is logical.

\paragraph{Outline}
The main technical point of this work is the workflow for the controller design, which is introduced in Section 2. In Section 3, we will practically show an example to realise the controller design and verify the theoretical prediction of the controller by agent-based simulations. In Section 4, we summarise the results and point out the related concepts and further directions.

\section{Workflow for controller design}
     In this study, our control-by-design (mechanism design)
     approach comes from \textbf{single-input pole assignment}
     approach  for linear system in the modern control theory,
     which has being applied on control the dynamics structure  \cite{Y5C2022}.
\begin{figure}
\centering
\includegraphics[width=0.7\textwidth]{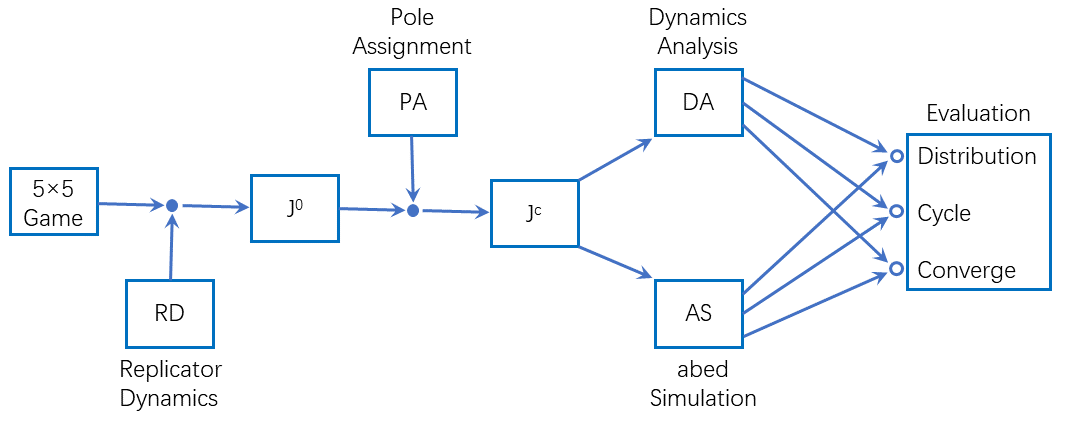}
\caption{Workflow:  1,$J^o$; 2, PA; 3, $J^C$; 4: DA; 5: Simulation; 6; Evaluation (a: Distribution, b: Cycle, and c: Convergence speed).
\label{fig:workflow2}}
\end{figure}
     As shown in Figure \ref{fig:workflow2}, the workflow includes follow steps:
    \begin{enumerate}
    \item Solve the dynamics equation for the original
    eigen-system including Jacobian ${J^o}$, eigenvalue $\lambda^o$, and eigenvector $v^o$;
    \item Assign the desired pole $\lambda^c$;
    \item Solve the gain matrix ${K}$ for the given ${B}$ for the controlled Jacobian ${J^c}$ from the original Jacobian ${J^o}$ (see Eq. \ref{eq:jk1}  referring to \cite{Y5C2022}).
    \item This step has two parallel parts:
\begin{enumerate}
\item [4.1]
    Derive the theoretical dynamics analysis results on the observation set, $O^T$.
\item [4.2]
    Conduct agent-based simulation and report the observation set, $O^S$.
\end{enumerate}
At this step, it is necessary to clarify the measurements of the observation.
    \item Check the controller's valuation by comparing $O^T$ and $O^S$. That is  to answer
    whether theoretical expectations are archived,
    meanwhile, whether the constrain conditions are satisfied.
\end{enumerate}
We will show how to realise the control
     by 5-strategy  game practically in an example.

\section{An example}
\subsection{Step 0: Game and its equilibrium}
We employ a five strategy symmetry game to show how to
observe the eigensystem and how to control the dynamics structure in experiment. The game is an one population 5$\times$5 symmetric game. In strategy state space, its evolution trajectory is of a 5 dimensional trajectory. Table \ref{tab:gamemodel} shows the 5-strategy payoff matrix. The game has and only has two Nash equilibrium \cite{enumNash2010}.
The first Nash equilibrium, denoted as Nash\_1, is
\begin{equation}\label{eq:Nash_eq1}
\rho_{\text{Nash\_1}}  = \frac{1}{3}(1, 1, 1, 0, 0),
\end{equation} %
in which the game falls into a 3-dimensional subspace, appearing as a rock-paper-scissors game, and its expected payoff is 2/3.
Intuitively, there should exist persistent endogenous cycles
along $x_1 \rightarrow x_2 \rightarrow x_3  \rightarrow x_1 ...$ when Nash\_1 is selected.
The second, denoted as Nash\_2, is
\begin{equation}\label{eq:Nash_eq2}
\rho_{\text{Nash\_2}}  = \frac{1}{2}(0, 0, 0, 1, 1),
\end{equation}
in which the game falls into a 2-dimensional subspace,
appearing as an anti-coordination game, and its expected payoff is 1/2. Intuitively, there exist no persistent endogenous cycles when Nash\_2 is selected.
%

\subsection{Step 1: The original eigensystem}
To illustrate the dynamics controller,
we use the replicator dynamics  \cite{2011Sandholm},
which is the original velocity vector field of the game, as the original system.
\paragraph{Original dynamics system}
The dynamics system can be expressed as follow:
\begin{equation}\label{eq:repliequl}
\Dot{x}_j=x_j\Big(U_j - {\overline{U}}\Big),
\end{equation}
in which $x_j$ is the $j$th strategy player's probability
in the population with the $j$th strategy player included, and
$\Dot{x}_j$ is the evolution velocity of the probability;
 $U_j$ the payoff of the $j$th strategy player;
And $overline{U}_X$ is the population's average payoff, which equals  $\sum_{k=1}^{5} x_k U_k$. This 5-dimensional space has one concentration, $\sum_j x_i=1 \cap x_i \geq 0 ~ (i \in \{1,2,...,5\})$, implying that the system must be in a 5-dimensional simplex space at all times. The Nash\_1 and Nash\_2 are the original velocity field's rest points.

\paragraph{Eigen system} The Jacobian matrix ${J^o}$ evaluated at
 ${\text{Nash\_1}}$ is
     $$
\small
J^o_{\text{Nash\_1}} =   \left[
     	\begin{array}{rrrrr}
     	 -\frac{4}{9} &-\frac{4}{9} &\frac{2}{9} & \frac{1}{9} & \frac{1}{9} \\
     	 \frac{2}{9} & -\frac{4}{9} & -\frac{4}{9} & -\frac{5}{9} & \frac{7}{9} \\
     	 -\frac{4}{9}& \frac{2}{9} & -\frac{4}{9} & \frac{7}{9} & \frac{4}{9} \\
     	 0& 0& 0& -1 & 0 \\
     	 0& 0& 0& 0 & -2 \\
     	\end{array}
     \right]. $$
And its eigenvalues $\lambda^o_{\text{Nash\_1}}$
are
\begin{equation}\label{eq:lambda_o}
    \lambda^o_{\text{Nash\_1}} = =\left[
     	\begin{array}{rrrrr}
     	-\frac{1}{3} + \frac{\sqrt{3}}{3}i & -\frac{1}{3} - \frac{\sqrt{3}}{3}i  &  -\frac{2}{3} & -1 & -2 \\
     	\end{array}
     \right].
\end{equation}
We choose the $\lambda^o_{\text{Nash\_1}}$ as the original pole in this study.
%
As all the real part of the eigenvalues are negative,
the system is stable at Nash\_1 locally.
The first two column of the eigenvector matrix is
the original complex eigenvector which  represents the dynamics structure. The third column represents the equilibrium distribution, because its related left eigenvector is the unit vector, [1 1 1 1 1], and its related eigenvalue is the oppose of the expected payoff at the equilibrium.
 In this example, the complex eigenvalue is the object of control-by-design.
In the same way, we can calculate the eigensystem at
$\rho_{\text{Nash\_2}}$. As results, 
the eigenvalues are
     $$\lambda^o_{\text{Nash\_2}} = \left[
     	\begin{array}{ccccc}
     	 -3/2 & -3/2 & -1/2 & -1/2 & 0 \\
     	\end{array}
     \right].$$
As all of the eigenvalues are negative real,
the system is also stable at Nash\_2 locally; Meanwhile, all the eigenvectors
are real, and no cyclic motion is expected around Nash\_2.
In general, each eqilibrium can be an object to control, but, in this study case, we do not try to control Nash\_2.

\subsection{Step 2: Assign control goal }\label{sec:assigngoal}

Our goal is to control
(plus or minus a small real number) the real parts of
the complex eigenvalues shown in Eq. \ref{eq:lambda_o}.
The assign pool can be  expressed as,
\begin{equation}\label{eq:design_goalm}
    ~{\lambda^c} =  ~{\lambda^o} + b~[0 ~0 ~0 ~1 ~1]
\end{equation}
Here, the $\lambda^o$ is the original pool at Nash\_1, and $\lambda^c$ is the controlled (optimal) pool.
That is, for each optimal goal $b$, in $\lambda^0$ of the original system matrix $J^o$, a pair of poles (eigenvalue) is to be shifted.
We set $b$ form $-1$ to 1 by step 0.2,
$b \in [-1, -0.8, -0.6, ..., 1]$, respectively.
Such control will change the stability and the attraction of the equilibrium Nash\_1.

As results, in theory, the equilibrium point Nash\_1
will be more attractive when $b \longrightarrow -1$, and Nash\_1 will be more likely being selected; alternatively,
the equilibrium point Nash\_1 will more
repelled when $b \longrightarrow +1$, and as a consequence, Nash\_2 will be more likely being selected.

\subsection{Step 3: Controller design}
The main philosophy of the controller design is an inverse solution procedure, which
recasts the controller design task to a
optimization problem with constraints \cite{pole2018}. In this study, we set two constraint conditions, equilibrium conservation and payoff conservation. The definition of the controller is shown in Section \ref{sec:cons_Jc}.

 Denote
$\mathbf{x}=[x_1 ~ x_2  ~ x_3 ~ x_4 ~ x_5 ]^T$,
referring to the definition of $J^c$, the controlled velocity field is
 \begin{eqnarray}
 \label{eq:jk1}
 \mathbf{\dot{x}} &=& {J^c} \mathbf{x} \nonumber \\
  &=& {(J^o + B K - \sum_{i=1}^{5}{k_i x_i})} \mathbf{x}.
 \end{eqnarray}
Notice that, as the relation --- $\sum_{i=1}^{5}{k_i x_i^*} = 0$ ($x^*$  is the Nash\_1) must hold. It is important, because the controller is required to preserve the  equilibrium.
The theory of optimal control is concerned with operating a dynamic system at minimum cost.
At this setting, the cost is zero.
Having the ${J^c}$ defined in Eq \ref{eq:jk1}, we can solve pole assignment problem, shown in Eq. \ref{eq:design_goalm}, for the gain matrix $K$.
%
In this study case,  we use the pole assignment algorithm
\cite{poleplacement1985} which is name as \textit{place}
function in Matlab as the tool to compute the matrices $K$ to
achieve the desired closed-loop pole locations.
The results are shown in Table \ref{tab:Kvalue}.

\begin{table}[!ht]
    \centering
    \begin{tabular}{c|r|r|r|r|r}
    \hline
~~~~$b$~~~~&$k_1$~~~~&$k_2$~~~~&$k_3$~~~~&$k_4$~~~~&$k_5$~~~~\\\hline
 $-$0.8&0.5247 & ~0.9485 & $-$1.4732 & $-$1.8335 & ~0.2335 \\
 $-$0.6&0.4843 & 0.5524 & $-$1.0368 & $-$1.3232 & 0.1232 \\
 $-$0.4&0.3834 & 0.2623 & $-$0.6458 & $-$0.8476 & 0.0476 \\
 $-$0.2&0.2220 & 0.0782 & $-$0.3002 & $-$0.4065 & 0.0065 \\
0& 0 & 0 & 0 & 0 & 0 \\
0.2& $-$0.2825 & 0.0277 & 0.2548 & 0.3719 & 0.0281 \\
0.4& $-$0.6256 & 0.1614 & 0.4641 & 0.7092 & 0.0908 \\
0.6& $-$1.0292 & 0.4011 & 0.6281 & 1.0119 & 0.1881 \\
 0.8&$-$1.4933 & 0.7467 & 0.7467 & 1.2800 & 0.3200 \\\hline
    \end{tabular}
    \caption{Controller gain matrix $K$'s value when ${B} = \big[0~ 0~ 0~ 1~ 1\big]^T.$}
    \label{tab:Kvalue}
\end{table}

\subsection{Step 4.1 Dynamics analysis results}\label{subsec:theoPred}
Now we need to evaluate the dynamics property of the controlled the velocity field.
Comparing to the original system, the controlled has observable consequence.
Due to the design aim, some of observation is invariant, and some changed.
The verifiable theoretical predictions of the controller are listed as following.

\begin{enumerate}
\item  \textbf{Distribution.}  by changing the
pole assignmant parameter $b$. The strategy distribution of
the controlled game should be identical to that of the original game.
The theoretical expectation is
\begin{eqnarray}
\rho^T\!&\rightarrow&  {\text{Nash\_1}} = \!\frac{1}{3}(1, 1, 1, 0, 0)~\text{~~~~when~}~ b  \rightarrow -1, \\
   \rho^T\!&\rightarrow&  {\text{Nash\_2}} = \!\frac{1}{2}(0, 0, 0, 1, 1)~\text{~~~~when~}~ b \!\rightarrow\!1.
\end{eqnarray}

Wherein the superscript indicates the signal of the parameter.
This is the prediction on equilibrium selection.

In measurement, the prediction of equilibrium section can be
verified by Euclidean distance ($d$) between $\rho^T$ and $\rho^E$, where $\rho^E$
is the strategy proposition vector in the time series from computer simulation or human experiment. In details,
  \begin{eqnarray}\label{eq:meanrho}
\bar{\rho}^E_{i} = \lim_{t \rightarrow \infty} \frac{1}{T}\sum_{t=0}^T \rho_{i}(t),
\end{eqnarray}
where $\rho_i(t)$ is
the proportion of $i$-th strategy used at time $t \in [0, T]$ . We can measure the time dependent
Euclidean distance $d(t)$ from Nash\_1 and
 Nash\_2 respectively,
\begin{eqnarray}
    d_{\text{Nash\_1}}(t) &=& \big|\rho(t) -  {\text{Nash\_1}}\big|   \\
    d_{\text{Nash\_2}}(t) &=& \big|\rho(t) -  {\text{Nash\_2}}\big|.  
\end{eqnarray}
We will see that,
\begin{eqnarray}
    d_{\text{Nash\_1}}(t) &=& \longrightarrow  0 ~\text{~~~~when~}~ b  \rightarrow -1, \\
    d_{\text{Nash\_2}}(t) &=& \longrightarrow  0 ~\text{~~~~when~}~ b  \rightarrow  1.
\end{eqnarray}

\item \textbf{Converge speed.} The converge speed to desired equilibrium is impacted by the pole assignment parameter, $b$.
%
The theoretical prediction is that,
disregarding which equilibrium selected, convergence speed will be faster when $|b|$ increase.

In measurement, the convergence speed is defined as the time cost of
time (denote as $\tau_{1/2}$) when $d(\tau_{1/2}) = d^0/2$,
in which $d_0$ is the Euclidean distance from  the full randomly
initial distribution [1, 1, 1, 1, 1]/5 to equilibrium
(Nash\_1 or Nash\_2). In details, $d^0_{\text{Nash\_1}}$ = 0.184 and $d^0_{\text{Nash\_2}}$ = 0.273. So, $\tau_{1/2}(b)$ can be obtained in the time series.

\item \textbf{Cycle.} In theory, the eigencycles
in the 2-d subspace of the game can be calculated from
the complex eigenvector associated to the eigenvalue shown in Eq. \ref{eq:lambda_o},
 referring to \cite{wang2022shujie}.
This can be verified by  time average of the angular momentum
measurement along the time series, referring to \cite{wang2022shujie},
\begin{eqnarray}\label{eq:meanL}
\bar{L}_{mn}&= & \frac{1}{t'}\sum_{t=0}^{t'} x_{mn}(t) \times x_{mn}(t + 1)
\end{eqnarray}
Herein, $x_{mn}(t)$ is the strategy vector in the $mn$ subspace
(2-d subspace), $t'$ is the length of time series.
For a 5 strategy game, the identical 2-d subspace number is 10,
and the observer sample is 10 \cite{wang2022shujie,2021Qinmei,WY2020}.
The equivalent of the theoretical eigencycle and the observed angular momentum $L$ in is proved referring to
\cite{2021Qinmei,WY2020}.

In measurement, the cycle strength $|\bar{L}|$ is defined as
\begin{equation}
    |\bar{L}| = \big(\sum_{mn} L_{mn}^2\big)^{1/2}
\end{equation}
in the time series to verify the theoretical expectation of the strength of the cycles.

\end{enumerate}

\subsection{Step 4.2: ABED Simulation}

Agent-based evolutionary dynamics (abed) simulation (see appendix \ref{app:agent}) are carried out for various control parameter $b$. Having the time series data from the abed simulation, we can evaluate whether
the theoretical expectation is supported by data.

\subsection{Step 5: Evaluate the controller}

\subsubsection{Statistical results}

The three figures associate to the
three theoretical predictions are shown, respectively.

\begin{figure}
\centering
\includegraphics[width=.55\textwidth]{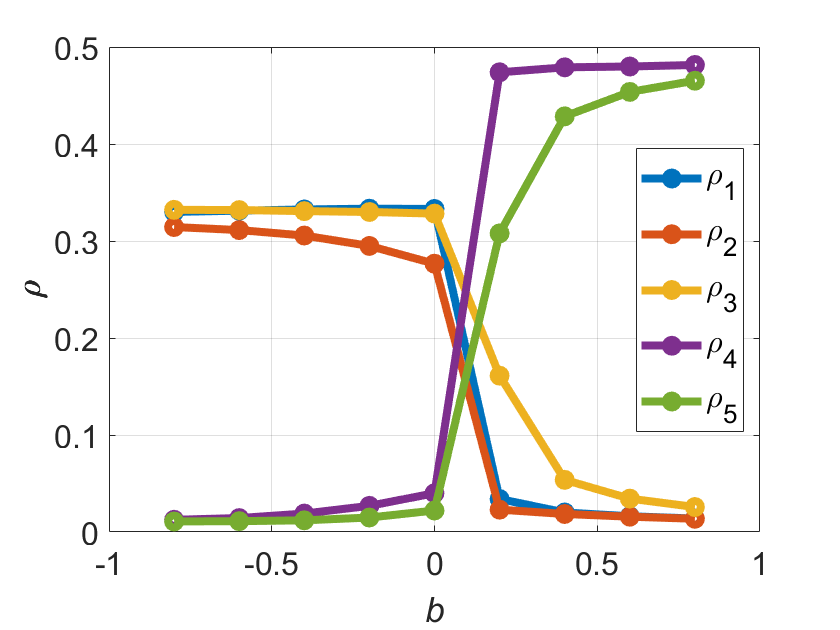}
 \includegraphics[width=.55\textwidth]{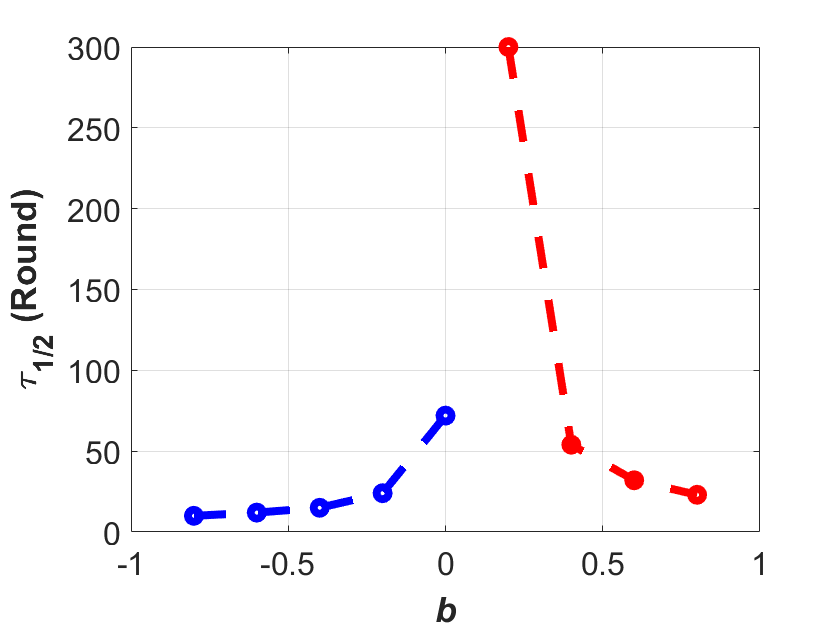}
 \includegraphics[width=.55\textwidth]{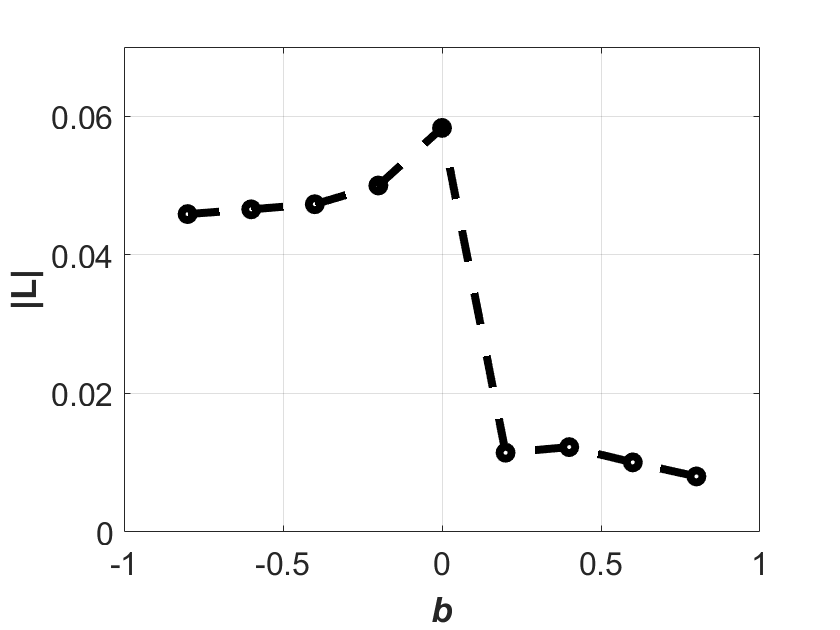}
\caption{\label{fig:distri_result}
(\textbf{a}) Distribution of the long run average
$\rho^S(b)$.
when $b  \rightarrow -1$,
the trend is Nash\_1 selected;
alternatively, when $b \rightarrow 1$,
 Nash\_2 selected.
 (\textbf{b}) Speed of convergence depends $b$.
When $b \rightarrow -1$, the convergence to Nash\_1 is faster. Alternatively, when $b \rightarrow 1$, the convergence to Nash\_2 is faster
 (\textbf{c}) cycle strength $L(b)$. }
\end{figure}

\begin{enumerate}
\item Figure \ref{fig:distri_result}(a) shows the distribution of the long run average
($\bar{\rho_1},\bar{\rho_2},
\bar{\rho_3},\bar{\rho_4},\bar{\rho_5}$)
as the function of the pole assignment parameter $b$.
As expected,
when $b$ being $ 0 \rightarrow -1$,
the trend is to select
 equilibrium Nash\_1;
Alternatively, when $b$ being $ 0 \rightarrow 1$,
the trend is to select equilibrium Nash\_2.
\item
Figure \ref{fig:distri_result}(b) shows the changing of the convergence time depends on the changing of the  parameter $b$. Convergence speed depends on pole assignment parameter $b$.
When $b \rightarrow -1$, observed $\tau_{1/2} \rightarrow 0$, which means the convergence to Nash\_1 is faster. Alternatively, when $b \rightarrow 1$, observed $\tau_{1/2} \rightarrow 0$ too, the convergence to Nash\_2 is faster.
\item
Figure \ref{fig:distri_result}(c) shows the eigencycle set by
changing the parameter $b$. Each curve presents an observation of $L_{mn}$ of the 10 eigencycle, referring to the pole assignment
parameter $b$ shown in Eq. \ref{eq:design_goalm}.
It is obviously that, when the equilibrium remained
at Nash\_1 the cycle in the $(x_1, x_2, x_3)$ subspace is significant;
Alternatively, when the equilibrium shift to the Nash\_2 equilibrium, the cycles disappear.
\end{enumerate}

\subsubsection{Explanation for the statistical result}
Referring to the statistical results, we can reach the following conclusion:
\begin{enumerate}
\item By pole assignment, the equilibrium section is realised as expect.
\item As a association phenomena,the measured cycle exists or
not meet the theoretical expectation well.
\item As a association phenomena, the measured cycle exists or
not meet the theoretical expectation well.
\end{enumerate}

 Base on these results, we can say, in a game with multi equilibrium,
 pole assignment can be apply to influence the equilibrium selection.
 At the same time, the associated observation relates to the equilibrium
 selection by pole assignment meet the general dynamics theory well.

 Importantly, our results suggest that, game dynamics system is
 not an exception of the long existed dynamics system
 which has been extensively studied and applied in engineering.
 Meanwhile it is not an expectation of the long existed modern control theory.

\section{Discussion}
Equilibrium selection is a critical concern in game theory. This study demonstrates a method of controlling equilibrium selection using a dynamic control approach. Through a straightforward example game and a workflow of state feedback control, the study shows that it is possible to control equilibrium selection and achieve a desired outcome. This is the first instance, to the best of our knowledge, where equilibrium selection has been controlled using the controllability of the velocity field.

\subsection{Related works on the control the game dynamics}
In recent years, there has been extensive research on game-environment interaction, led by \cite{2016feedback}. One approach is to incorporate the feedback mechanism from the environment into replicator dynamics, whereby the feedback changes the pay-off structure and influences the evolution of strategies. Several studies have explored this idea, such as \cite{2016feedback,2015evolutionary,2020FuFeng}.

Our method distinguishes itself from the conventional approach of treating game-environment feedback as state-dependent feedback within the game dynamics process. Instead, our feedback controller utilizes pole assignment techniques based on the eigensystem, which is grounded in modern control theory. Essentially, the core of our design strategy is to regulate the velocity field, which in turn governs the selection of equilibrium.

\subsection{Further directions of the controller design}
We have noticed the two limitation of the approach for equilibrium selection as follow.
\begin{enumerate}
\item The general condition for the approach being applied for equilibrium selection is not know. In this study, we have show only a special game case \ref{tab:gamemodel} to illustrate
the ability of pole assignment for equilibrium selection.
In this case, the left-up 3 by 3 related to rock-paper scissors,
and the right-down 2 by 2 is an anti coordination game, and is a special case.
Referring to \cite{morse1970}, this question is about pole assignability in game dynamics.

\item For various game dynamics model, whether there is general method
to design the equilibrium selection is not known.
The algorithm designed for the controller shown in Eq. \ref{eq:BKT},
especially the third term (the financial balance term $T$) is specified for
the replicator dynamics shown in Eq. (\ref{eq:rp}).
\end{enumerate}
It is important to emphasize the need for careful evaluation of the consequences of the control-by-design approach (as demonstrated in Equation \ref{eq:jk1}) when using it for equilibrium selection. We also hope to see the development of more general solutions in the future.
%

%
%
%
%
%
%
%
%
%
%
%
%
%
%
%
%
%
%
%
%
%
%

 \bibliographystyle{plain}

%
%
%
%
%
%
%
%
%
%
%
%
%
%
%
%
%
%
%
%
%
%

\clearpage
\section{Appendix}

\subsection{Pole assignment}
Pole assignment  (called also as pole placement)
is a key approach in modern control theory.
It is design methodology wherein the objective is to
place the eigenvalues of the closed-loop system in
desired regions of the complex plane \cite{dorf1995modern}.
It is a method employed in feedback control system theory
to place the closed-loop poles (or eigenvalues)
of a plant in desired locations.
Placing poles is desirable because the location of
the poles corresponds directly to the eigenvalues of the system,
which control the characteristics of the response of the system.
After more than 50 years study,
the pole assignment has becomes basic approach
in modern control theory textbook in engineering field \cite{pole1989,pole1998book,pole2017modern}.

It is not surprise that the pole assignment is a important role
when we facing dynamics system. This is due to
our knowledge system to handle complex  dynamics system.
It is well known that \cite{strogatz1994nonlinear},
in dynamical systems, in general, do not have closed-form solutions,
and its behavior is hard to be predicted and controlled.
However, linear dynamical systems can be solved exactly.
Linear dynamical systems have a rich set of mathematical properties \cite{kalman1963mathematical}.
Linear systems can also be used to understand
the qualitative behavior of general dynamical systems,
by calculating the \textit{equilibrium points} of
the system and approximating it as
a linear dynamics behavior around each such points.
The central mathematical concept set around
the equilibrium is the \textit{eigen system}.
The concept set includes eigenvalue, eigenvector,
eigen space, invariant manifold and so on. In eigen system,
eigenvalue (pole) play the cruel role because
the pole determines the stability of the equilibrium.
So, if we are able to assign the pole in theory,
we are able to control the stability of the equilibrium.
This is the logic line explaining why pole assignment
is important for dynamics system.

It is not surprise, too, that the pole placement approach
has been extensively applied in real system since 1970s.
Examples are to helicopter stabilization systems
\cite{pole1973application}, to the spacecraft motion control
\cite{zubov2013modification}, to electro-hydraulic servosystem design  \cite{plummer1997decoupling},
to active control of car longitudinal oscillations
\cite{richard1999polynomial}. Game theory study
multi agent strategy interactions and has widely application.
However, surprisingly, the pole assignment approach
has rare appear in the field covered by game theory.

 %
 %
 %

\subsection{Algorithm for the controller design}
\subsubsection{Basic} Assume that, an one population $n$ strategy game
whose evolutionary dynamics can be described by the replicator dynamics,
\begin{equation}\label{eq:rp}
\dot{x_i} = F(x).
\end{equation}
This is a time-invariant systems, because time $t$  does not appear
as a variable in the rhs terms. Assume again that,
the game system can be linearized and expanded as
\begin{equation}\label{eq:jo}
\dot{x}={J^o} x,
\end{equation}
in which, ${J^o}$ is the Jacobian matrix at  equilibrium (rest point), and $x$ is a small deviation from the equilibrium. If $J^o$ is diagonalizable, we call its eigenvalue set as original pole vector, $\lambda^o$.
In mathematics, the game dynamics is specified on the $n$-simplex by
an ordinary differential equation (ODE). In control engineering,
this is a continuous-time LTI   (linear time-invariant) system,
which is a fairly common situation to carry out feedback control
(see wiki: State-space\_representation).

The original pole assignment approach states that, given the single- or multi-input system
\begin{equation}\label{eq:oriPole}
    \dot{x} = J^o x + B u
\end{equation}
and a vector $\lambda^c$ of desired self-conjugate closed-loop pole locations, place computes a \textbf{gain matrix $K$} such that the state feedback $u = -Kx$ places the closed-loop poles at the locations $\lambda^c$. In other words, the eigenvalues of $J^o - BK$ match the entries of $\lambda^c$.

Admittedly, the original pole assignment approach does not fit the game dynamics system. Because, for a given $B$, for example $B=[0~0~0~0~1]$, once $Bu$ term is not zero, the sum of rhs of Eq. (\ref{eq:oriPole}) is not zero. This deviates from the simplex concentration, $\sum_i x_i = 1$. So, the original approach has to be improved.

\subsubsection{Construct ${J^c}$}\label{sec:cons_Jc}
We use the linear control approach, pole assignment, to design the control term. The control terms includes \textbf{gain matrix} $K$ control vector (an $1 \times n$ vector, a row vector) on the \textbf{channel $B$} (a $n \times 1$ vector, a column vector), and a \textbf{tax term $T$} for the payoff conservation. The payoff's conservation can be understood as budget balance. Then, the controlled velocity field can be expressed as
    \begin{equation}\label{eq:BKT}
          \dot{x}={J^o} \cdot x + {B} \cdot {K} \cdot x + T \cdot x
    \end{equation}
Here, $B \cdot K$ is the ordinary linear state dependent control term
(see standard text book of modern control system), which is called as \textbf{reward term}.
The financial balance term
\begin{equation}\label{eq:T}
    T := - \sum_{i=1}^n B_i \cdot \sum_{j=1}^n K_j  x_j
\end{equation}
is a constant value defined at each given state $x$. It can be verified that, at each period in the game, each agent is taxed the constant value, $T$. The tax term makes the controller to be budget balanced—the total amount of tax collected from the system at a given period is exactly equal to the total amount of reward given to the system at that time.

The controlled Jacobian ${J^c}$ can be expressed as
\begin{equation}\label{eq:jc1}
{J^c} = {J^o} + {B} \cdot {K} + T.
\end{equation}
Then for an assigned pool $\lambda^c$ and a given $B$, we can obtain $K$. Once ${K}$ is given , ${J^c}$
as well as its the eigen system is determined.

\paragraph{On Constraint condition of the controller design}
\begin{enumerate}
\item  \textbf{Equilibrium conservation:} The equilibrium (rest point)  should
be preserved and not be shifted; This can be satisfied by setting $K \cdot x_{\text{Nash}} = 0$ when solving the character equations for $J^c$.
\item
The controller, $J^c$, does not change the expected payoff. Or saying, the equilibrium payoff remains an invariant by $J^c$. \\
\textbf{Proof}:
Denote
    $H_0=\sum_{j=1}^{n}k_j x_j$,
referring to the definition of $T$ in Eq. (\ref{eq:T}), we have
\begin{equation}
    T=-H_0\sum_{m=1}^n B_m,
\end{equation}
which is a scalar at given $x$. Recall Eq. (\ref{eq:BKT}), applying the definition of
a element of Jacobian $J_{ij} = \partial \dot{x_i} / \partial x_j$, we have
\begin{equation}
    J_{ij}^c=J_{ij}^0+B_i\cdot k_j
    - k_j   \sum_{m=1}^n B_m \cdot x_i
\end{equation}
Sum over each column, we can reach that,
\begin{eqnarray}
            \sum_{i=1}^n J_{ij}^c &= &\sum_{i=1}^n J_{ij}^0
    +\sum_{i=1}^n B_i   k_j
    -  k_j   \sum_{m=1}^n B_m \sum_{i=1}^n x_i \\
    &=& \sum_{i=1}^n J_{ij}^0
\end{eqnarray}
 It is known that \cite{dan2016}, there exists a left
 eigenvector (1, 1, 1, ..., 1) for $J$ and its related right eigenvector
 is the equilibrium distribution, and the sum of the column value of $J$
 is the equilibrium payoff. Having the controller,
 the system earning (expected payoff) is remained.
\item \textbf{Payoff conservation:} There is no additional payoff (financial support) adding to the game system during the controlling processes, which naturally can be realised by the definition of the tax term $T$ is of financial balance.

\begin{enumerate}
 \item $K$ rewards the strategy at the channel $j$ determined by $B$, of which the reward value is $({BKx})_j$.
As a result, for each agent using strategy $j$,
its reward (additional payoff) is $({BKx})_j / x_j$.
\item $T$ taxes each strategy equally. The tax value is a constant, $T$, depending on state $x$. For each individual agent
its tax value is $T$.
$T$ term balance the additional pay
due to the  gain matrix $K$ above to each agent in game.
\end{enumerate}

\end{enumerate}


\subsubsection{Explain $B$}

In this study, we limit ourselves to illustrate an example of
the equilibrium selection by a controller designed by pole assignment.
We choose ${B} =[0~ 0~ 0~ 1~ 1]^T$ for the illustration the controllablity. We hope to emphasise that, for a multi equilibrium system,
the choosing of $B$
relates to whether a controller having a solution for its desired goal. For game dynamics, the general answer on $B$ choosing  is out of the scope of this study.

\subsection{Agent-based evolutionary dynamics (ABED) simulation \label{app:agent}}
 The simulation is introduced as following.

\begin{enumerate}
 \item  Select simulation platform:
  We use ABED (Agent-based evolutionary dynamic) simulator \cite{2019abed}, which is widely used in
  the field to study evolutionary game dynamics.
  The platform has integrated various learning rules and matching rules.
  The platform is of agent-based long-running repeated game setting.  The simulator is implemented in the open-source platform NetLogo 6.0.4.
  \item Setting parameters: The simulation is under imitative protocols,
  in which candidates are agents; meanwhile,
  the decision method is of pairwise-comparison of the strategy payoff.
  The complete-matching is set. These setting are
  follow the user guide of the platform,
  which system will performs like replicator dynamics
  shown in Eq. (\ref{eq:rp}) in large population (1000 agents)
  and low reversion probability 5\%) limit.
 \item Add control modular to the platform:
  add a modular to control the agents payoff referring to
  the algorithm shown in Eq. (\ref{eq:BKT}).
 \item Specify controller parameter:
  the modular is a state-depend feedback control 'device'.
  The device is specified according to the parameter vector $B$ and $K$.
  Various parameter assign can archive various goal.
  In this case, $B$ and $K$ are shown in Table \ref{tab:Kvalue}.
 \item Conduct the simulation:
   In our study case, time cost for each simulation of a given
   parameter $b$, for 6000 rounds, is about 1 minute in a desktop personal computer,
   which CPU is 3.70 GHz and the memory is 8 GB.
 \item Analysis the time series:  Main outcome of the simulator
 is the time series
of the strategies density and related variables.
\end{enumerate}

\begin{table}
\caption{The parameter setting for the ABED simulations}
\label{tab:sim_para}
\begin{center}
\begin{tabular}{|r|r|}
   	 \hline
Parameter&	Replicator $[S_1]$ 		\\
   	 \hline
payoff-matrix  & [[ 0  0  2  0 -2]~ 		\\
               &  [ 2  0  0 -2  0]~ 	 	\\
               &  [ 0  2  0  2  -1]~  	\\
               &  [ -2  0  1  0  1]~	 	\\
               &  [0  -2  -2  1  0]]	 	\\
n-of-agents&	1000	\\
random-initial-condition?&	TRUE 		\\
initial-condition&	[200 200 200 200 200] 	\\
candidate-selection&	imitative 	\\
n-of-candidates&	2	\\
decision-method&	pairwise- 		\\
 &	difference 	\\
complete-matching?&	TRUE 	\\
n-of-trials&	999 	\\
single-sample?&	TRUE  	\\
tie-breaker&	uniform 		\\
log-noise-level&	0.031	\\
use-prob-revision?&	TRUE 	\\
prob-revision&	0.2	\\
n-of-revisions-per-tick&	10\\
prob-mutation&	0.05	\\
trials-with- replacement?&	FALSE 		\\
self-matching?&	FALSE 		\\
imitatees-with-replacement?&	FALSE 	\\
consider-imitating-self?&	FALSE  	\\
plot-every-?-secs&	0.2		\\
duration-of-recent&	10		\\
show-recent-history?&	TRUE 	\\
show-complete-history?&	TRUE	\\
 K  &	 [-1.49 0.75 0.75 1.28 0.32] 	\\
 B\_Channel & [0 0 0 1 1] 		\\
     	 \hline
\end{tabular}
\end{center}
\end{table}

\subsection{Abbreviations} \label{sec:app-abbreviations}
The abbreviations and mathematical symbols, shown in Table \ref{tab:app-abbr}, are used in the main text, this appendix and the supplementary information.
 \begin{table}[ht!]
    \centering
    \begin{tabular}{lll}
 \hline
 $b$   &&  pole assignment parameter, $b \in [-1,1]$ \\
$B$    && channel matrix, an $5 \times 1$ vector, a column vector \\
$K$    && gain matrix, an $1 \times 5$ vector, a row vector \\
 $B \cdot K$ & & Reward term, a matrix with real number elements.\\
       & & It is independent of the state vector. \\
 $T$     & &  Tax term, a constant, a real number. \\
       & & It is dependent of the state vector. \\
PTDE && Predictable temporary deviation from equilibrium\\
ABED && Agent-based evolutionary dynamics simulation \\
    \hline
    $\rho$ & & The proportion vector of strategies used \\
    &$\rho_i$&  The $i$-th component of $\rho$, \\
    &$\rho_i(t)$&   The $\rho_i$ observed at time (round) $t$ \\
    Nash\_1 & & The Nash equilibrium,  $\rho=(1/3,~1/3,~1/3,~0,~0)$  \\
    Nash\_2 & & The Nash equilibrium,  $\rho=(0,~0,~0,~1/2,~1/2)$ \\
     $d$ & & Euclidean distance between two distribution     \\
     && vectors, used to evaluate their difference. \\
      & $d_{\text{Nash\_1}}$& The $d$ between the observed distribution to Nash\_1     \\
      & $d_{\text{Nash\_2}}$& The $d$ between the observed distribution to Nash\_2     \\
     \hline
   $ o $&  & A fixed point, rest point, or zero velocity point, equilibrium   \\
   $\lambda$ & & Eigenvalue   \\
   &$\lambda^o$  &  Original pool \\
  & $\lambda^c$  &  Controlled pool, assign pole \\
   $\xi$ & & Eigenvector   \\
   $\eta_i$  & & The $i$-th component of a given eigenvector $\xi$  \\
    $L_{mn}$ &  & The eigencycle set in the 2D subspace set of a game  \\
    & & space, describing the cyclic  motion strength    \\
& $L_{(m,n)}$ & The $L$  at
       dimension (1,2):=($\rho_m,\rho_n$)  \\
 \hline
    \end{tabular}
    \caption{Abbreviations and mathematical symbols.}
    \label{tab:app-abbr}
\end{table}

\begin{figure}
\centering
 \includegraphics[width=.7\textwidth]{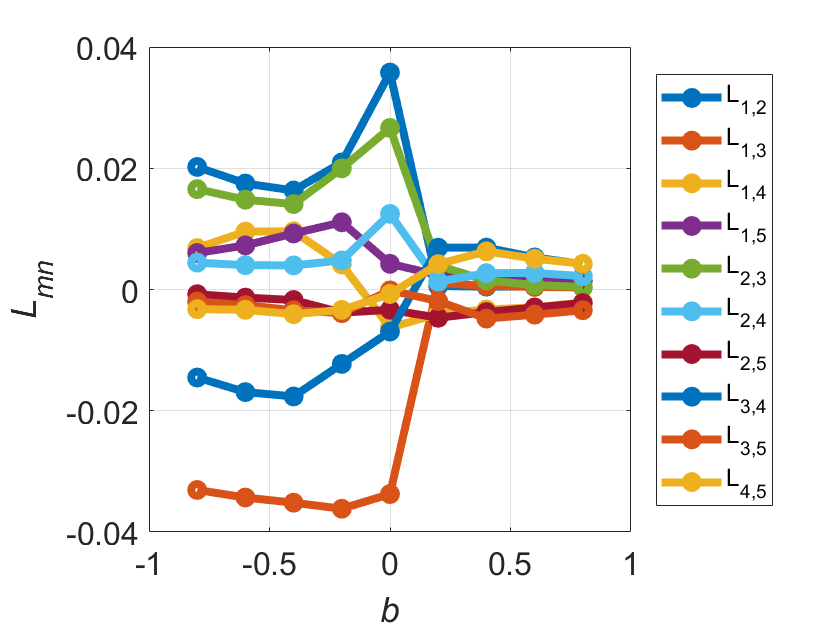}
\caption{\label{fig:cycle10}
Eigencycle strength $L_{mn}(b)$.}
\end{figure}

\begin{figure}
\centering
\includegraphics[width=1\textwidth,angle=0]{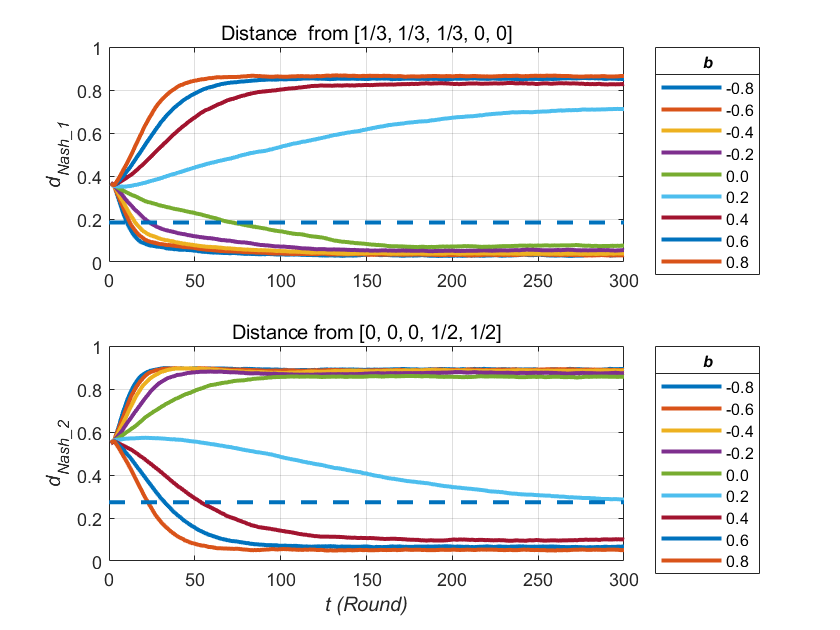}
\caption{\label{fig:d_t}
Speed of convergence depends on pole assignment parameter $b$.
When $b \rightarrow -1$, the convergence to Nash\_1 is faster. Alternatively, data shows that when $b \rightarrow 1$, the convergence to Nash\_2 is faster. }
\begin{verbatim}
K>> t_184(1:5)
ans =  10    12    15    24    72
K>> t_273(6:9)
ans =  300    54    32    23
\end{verbatim}
\end{figure}

\subsection{Data and code}\label{sec:data-code}

The documents containing the data and code are listed in Table 18 in SI and are accessible at http:github.com-xxxxxx-yyyyyy.

\clearpage
\begin{table}[ht!]
    \centering
    \begin{tabular}{l|l|l}
        \hline
        file name & ~~~~~~~~~~~~~description ~~~~~~~~~~~~~~~~&  ~~~~ note \\
        \hline
         E5C.nlogo &  ABED simulator code & Simulation \\
        \hline
        E5Cdata.rar & data generated by E5C.nlogo  &   \\
        &  x\_parameter.csv parameter & \\
        &  x\_A\_s.csv strateg& \\
        &  x\_A\_w.csv payoff & \\
        \hline
         T24simAll.m & mean $\rho$  &   \\
          &  mean $L_{mn}(b)$  &   \\
          & $d_{\text{Nash\_}i}(t)$ & \\
        \hline
        SLXMAT2.mat & data generated by T24simAll.m &   \\
                    & used by T24sim3Fig.m &   \\
       \hline
          T24sim3Fig.m & Fig 3(a)  & $\bar{\rho}_i(b)$ \\
          & Fig 3(b)  & $\tau_{1/2}$ \\
          & Fig 3(c)  &  $|L|$  \\
          & Fig 4  &  $L_{m,n}$  \\
          & Fig 5(a)  &  $d_{\text{Nash\_1}}(t)$  \\
          & Fig 5(b)  &  $d_{\text{Nash\_2}}(t)$  \\
       \hline
       read\_5angl\_forN3E.m & function called by T24simAll.m &  \\
        \hline
    \end{tabular}
    \caption{Data and code}
    \label{tab:da_code}
\end{table}

 
\end{document}